# Syndrome-Coupled Rate-Compatible Error-Correcting Codes

**Pengfei Huang**[*], **Yi Liu**[*], **Xiaojie Zhang**[†], **Paul H. Siegel**[*], and **Erich F. Haratsch**[‡]

[*]Electrical and Computer Engineering Dept., University of California, San Diego, La Jolla, CA 92093 U.S.A

[†]CNEX Labs, San Jose, CA 95128 U.S.A

[‡]Seagate Technology, Fremont, CA 94538 U.S.A

{pehuang,yil333,psiegel}@ucsd.edu

*Abstract*—Rate-compatible error-correcting codes (ECCs), which consist of a set of extended codes, are of practical interest in both wireless communications and data storage. In this work, we first study the lower bounds for rate-compatible ECCs, thus proving the existence of good rate-compatible codes. Then, we propose a general framework for constructing rate-compatible ECCs based on cosets and syndromes of a set of nested linear codes. We evaluate our construction from two points of view. From a combinatorial perspective, we show that we can construct rate-compatible codes with increasing minimum distances. From a probabilistic point of view, we prove that we are able to construct capacity-achieving rate-compatible codes.

## I. Introduction

Rate-compatible error-correcting codes (ECCs) consist of a set of extended codes, where all symbols of the higher rate code are part of the lower rate code. This allows to match the code rate of the sent data to the channel conditions by retransmitting incremental redundancy to the receiver. Such a scheme is known as hybrid automatic repeat request (HARQ) in wireless communications [13].

The idea of rate-compatible codes dates back to Davida and Reddy [3]. The most commonly used way to construct such codes is by puncturing; that is, to start with a good low-rate code and then successively discard some of the coded symbols (parity-check symbols) to produce higher-rate codes. This approach has been used for algebraic codes [3] [20], convolutional codes [7], turbo codes [17] [14], and low-density parity-check (LDPC) codes [6] [4]. The performance of punctured codes depends on the selected puncturing pattern. However, in general, determining good puncturing patterns is non-trivial, usually done with the aid of computer simulations.

The second approach is by extending; that is, to start with a good high-rate code and then successively add more parity-check symbols to generate lower-rate codes. A two-level extending method called Construction X was introduced in [15], and later was generalized to Construction XX [1]. Both constructions were utilized to find new codes with good minimum distance. In [10], codes from the extending scheme were used for HARQ systems. Extension-based rate-compatible LDPC codes were designed in [19] [12]. More recently, the extending approach was used to construct capacity-achieving rate-compatible polar codes [11] [8].

The goal of this paper is to provide a systematic approach for constructing rate-compatible codes with theoretically guaranteed properties. We use the extending approach and propose a new algebraic construction for rate-compatible codes; the properties of constructed codes are then analyzed from both combinatorial and probabilistic perspectives. Our contributions are as follows: 1) lower bounds are derived for rate-compatible codes, which have not been fully explored before; 2) a simple and general construction based on cosets and syndromes is proposed to construct rate-compatible codes, and some examples are given; 3) minimum distances of the constructed codes are determined, decoding algorithms are presented, and correctable error-erasure patterns are studied; 4) a connection to recent capacity-achieving rate-compatible polar codes is made.

The remainder of the paper is organized as follows. In Section II, we give the formal definition of rate-compatible codes and introduce notation used in the paper. In Section III, we study lower bounds for rate-compatible codes. In Section IV, we present a general construction for $M$-level rate-compatible codes, whose minimum distances are studied. Correctable patterns of errors and erasures are also investigated. In Section V, we show our construction can generate capacity-achieving rate-compatible codes by choosing the component codes properly. Section VI concludes the paper.

## II. Definitions and Preliminaries

In this section, we give the basic definitions and preliminaries that will be used in the paper.

We use the notation $[n]$ to denote the set $\{1,\ldots,n\}$. For a length-$n$ vector $\boldsymbol{v}$ over $\mathbb{F}_q$ and a set $\mathcal{I}\subseteq [n]$, the operation $\pi_\mathcal{I}(\boldsymbol{v})$ denotes the restriction of the vector $\boldsymbol{v}$ to coordinates in the set $\mathcal{I}$, and $w_q(\boldsymbol{v})$ represents the Hamming weight of the vector over $\mathbb{F}_q$. For two vectors $\boldsymbol{v}$ and $\boldsymbol{u}$ over $\mathbb{F}_q$, we use $d_q(\boldsymbol{v},\boldsymbol{u})$ to denote their Hamming distance. The transpose of a matrix $H$ is written as $H^T$. A linear code $\mathcal{C}$ over $\mathbb{F}_q$ of length $n$, dimension $k$, and minimum distance $d$ will be denoted by $\mathcal{C}=[n,k,d]_q$ or by $[n,k,d]_q$ for simplicity; in some cases, we will use notation $[n,k]_q$ to indicate only length and dimension. For any integers $a>b$, the summation in the form of $\sum_{i=a}^{b} X_i$ is defined to be 0. A binomial coefficient $\binom{a}{b}$ is defined to be 0 if $a<b$. For a set $\mathcal{C}$, $|\mathcal{C}|$ represents its cardinality. The $q$-ary entropy function $H_q: [0,1]\to[0,1]$, is defined by $H_q(x)=-x\log_q x-(1-x)\log_q(1-x)+x\log_q(q-1)$.

Now, we present the definition of rate-compatible codes.

**Definition 1.** *For $1\leqslant i\leqslant M$, let $\mathcal{C}_i$ be an $[n_i,k,d_i]_q$ linear code, where $n_1<n_2<\cdots<n_M$. The encoder of $\mathcal{C}_i$ is denoted by $\mathcal{E}_{\mathcal{C}_i}: \mathbb{F}_q^k \to \mathcal{C}_i$. These $M$ linear codes are said to be $M$-level rate-compatible, if for each $i, 1\leqslant i\leqslant M-1$, the following condition is satisfied for every possible input $\boldsymbol{u}\in\mathbb{F}_q^k$,*

$$\mathcal{E}_{\mathcal{C}_i}(\boldsymbol{u}) = \pi_{[n_i]}\big(\mathcal{E}_{\mathcal{C}_{i+1}}(\boldsymbol{u})\big). \qquad (1)$$

*We denote this $M$-level rate-compatible relation among these codes by $\mathcal{C}_1\prec\mathcal{C}_2\prec\cdots\prec\mathcal{C}_M$.*

**Remark 1.** For $1 \leq i \leq M - 1$, the rates satisfy $R_i = \frac{k}{n_i} > R_{i+1} = \frac{k}{n_{i+1}}$, but the minimum distances obey $d_i \leq d_{i+1}$. For systematic codes, the condition in (1) indicates that the set of parity-check symbols of a higher rate code is a subset of the parity-check symbols of a lower rate code.

In this paper, we will use the memoryless $q$-ary symmetric channel $W$ with crossover probability $p$. For every pair of a sent symbol $x \in \mathbb{F}_q$ and a received symbol $y \in \mathbb{F}_q$, the conditional probability is:
$$\Pr\{y|x\} = \begin{cases} 1 - p & \text{if } y = x \\ p/(q-1) & \text{if } y \neq x \end{cases}$$

The capacity of this channel is $C(W) = 1 - H_q(p)$ [16].

For a linear code $\mathcal{C} = [n, k, d]_q$ over a $q$-ary symmetric channel, let $P_e^{(n)}(x)$ denote the conditional block probability of error, assuming that $x$ was sent, $x \in \mathcal{C}$. Let $P_e^{(n)}(\mathcal{C})$ denote the average probability of error of this code. Due to symmetry, assuming equiprobable codewords, it is clear that,
$$P_e^{(n)}(\mathcal{C}) = \frac{1}{|\mathcal{C}|} \sum_{x \in \mathcal{C}} P_e^{(n)}(x) = P_e^{(n)}(x).$$

## III. LOWER BOUNDS FOR RATE-COMPATIBLE CODES

In this section, we derive lower bounds for rate-compatible codes.

### A. A General Lower Bound for M-Level Rate-Compatible Codes

Based on the technique used in the derivation of the Gilbert-Varshamov (GV) bound, we derive a GV-like lower bound for $M$-level rate-compatible codes.

**Theorem 2.** *There exist $M$-level rate-compatible codes $\mathcal{C}_1 \prec \mathcal{C}_2 \prec \cdots \prec \mathcal{C}_M$, where $\mathcal{C}_i = [n_i = n_1 + \sum_{j=2}^{i} r_j, k, \geq d_i]_q$ for $1 \leq i \leq M$, if the following inequalities are satisfied for all $1 \leq i \leq M$,*
$$d_i = \max\left\{d: \sum_{m=0}^{d-2} \binom{n_1 + \sum_{j=2}^{i} r_j - 1}{m}(q-1)^m < \frac{q^{n_1 + \sum_{j=2}^{i} r_j - k}}{M}\right\}. \tag{2}$$

*Proof:* The proof is based on a combinatorial argument. See Appendix A. ∎

The following corollary follows from Theorem 2, which shows that there exist good rate-compatible codes in the sense that each code can meet the corresponding asymptotic GV-bound.

**Corollary 3.** *There exist $M$-level rate-compatible codes $\mathcal{C}_1 \prec \mathcal{C}_2 \prec \cdots \prec \mathcal{C}_M$, where $\mathcal{C}_i = [n_i, k = R_i n_i, \geq \delta_i n_i]_q$ for $1 \leq i \leq M$ and $\delta_M \leq 1 - (1/q)$, simultaneously meeting the asymptotic GV bound:*
$$R_i \geq 1 - H_q(\delta_i). \tag{3}$$

*Proof:* Let $V_q(n, t) = \sum_{m=0}^{t} \binom{n}{m}(q-1)^m$. From Theorem 2, there exist $M$-level rate-compatible codes $\mathcal{C}_i = [n_i, k = R_i n_i, \geq \delta_i n_i]_q$ for $1 \leq i \leq M$ such that
$$V_q(n_i - 1, \delta_i n_i - 1) \geq \frac{q^{n_i - k}}{M}. \tag{4}$$

Since $V_q(n, t) \leq q^{n H_q(t/n)}$ for $0 \leq t/n \leq 1 - (1/q)$ [16], we have
$$q^{n_i H_q(\delta_i)} \geq V_q(n_i, \delta_i n_i) \geq V_q(n_i - 1, \delta_i n_i - 1) \geq \frac{q^{n_i - k}}{M},$$
which gives $R_i \geq 1 - H_q(\delta_i) - \frac{\log_q M}{n_i}$. As $n_i$ goes to infinity, we obtain the result. ∎

### B. A Lower Bound for Two-Level Rate-Compatible Codes with Known Weight Enumerator

For two-level rate-compatible codes, if the weight enumerator of the higher rate code is known, we have the following lower bound.

**Theorem 4.** *Let $\mathcal{C}_1$ be an $[n_1, k, d_1]_q$ code with weight enumerator $A(s) = \sum_{w=0}^{n_1} A_w s^w$, where $A_w$ is the number of codewords of Hamming weight $w$. There exist two-level rate-compatible codes $\mathcal{C}_1 \prec \mathcal{C}_2 = [n_2 = n_1 + r_2, k, \geq d_2]_q$, if*
$$\sum_{w=1}^{d_2 - 1} B_w < q^{r_2},$$
*where $B_w = \frac{1}{q-1} \sum_{m=1}^{w} A_m \binom{r_2}{w-m}(q-1)^{w-m}$, for $1 \leq w \leq n_2$.*

*Proof:* The proof is based on a probabilistic argument. See Appendix B. ∎

**Example 1.** Let code $\mathcal{C}_1$ be an $[n_1 = 15, k = 11, d_1 = 3]_2$ Hamming code, whose first few terms of the weight enumerator are: $A_0 = 1$, $A_3 = 35$, and $A_4 = 105$. Setting $r_2 = 11$, we have $\frac{\sum_{w=1}^{4} B_w}{2^{r_2}} = \frac{A_3 + A_3 \binom{r_2}{1} + A_4}{2^{r_2}} < 1$. From Theorem 4, there is a code $\mathcal{C}_2 = [n_2 = 26, k = 11, \geq 5]_2$ such that $\mathcal{C}_1 \prec \mathcal{C}_2$. □

## IV. A GENERAL CONSTRUCTION FOR M-LEVEL RATE-COMPATIBLE CODES

In this section, we present a general construction for $M$-level rate-compatible codes $\mathcal{C}_1 \prec \mathcal{C}_2 \prec \cdots \prec \mathcal{C}_M$. We then derive their minimum distances. The decoding algorithm and correctable error-erasure patterns are studied. We focus on the combinatorial property here and will leave the discussion on the capacity-achieving property of our construction to the next section.

In our construction for $M$-level rate-compatible codes, we need a set of component codes which are defined as follows.

**1)** Choose a set of nested codes $\mathcal{C}_1^M \subset \mathcal{C}_1^{M-1} \subset \cdots \subset \mathcal{C}_1^1 = \mathcal{C}_1 = [n_1, k, d_1]_q$, where $\mathcal{C}_1^i = [n_1, n_1 - \sum_{m=1}^{i} v_m, d_i]_q$ for $1 \leq i \leq M$. We have $k = n_1 - v_1$ and $d_1 \leq d_2 \leq \cdots \leq d_M$. Define $\mathcal{C}_1^0 = \emptyset$ and for $1 \leq \ell \leq i$, let matrix $H_{\mathcal{C}_1^\ell \setminus \mathcal{C}_1^{\ell-1}}$ represent a $v_\ell \times n_1$ matrix over $\mathbb{F}_q$ such that $\mathcal{C}_1^i$ has the following parity-check matrix:
$$H_{\mathcal{C}_1^i} = \begin{bmatrix} H_{\mathcal{C}_1^1} \\ H_{\mathcal{C}_1^2 \setminus \mathcal{C}_1^1} \\ \vdots \\ H_{\mathcal{C}_1^i \setminus \mathcal{C}_1^{i-1}} \end{bmatrix}.$$

The encoder of code $\mathcal{C}_1$ is denoted by $\mathcal{E}_{\mathcal{C}_1} : \mathbb{F}_q^k \to \mathcal{C}_1$. We also use $\mathcal{E}_{\mathcal{C}_1}^{-1}$ as the inverse of the encoding mapping

2) For $i$th level, $2 \leqslant i \leqslant M$, consider a set of auxiliary nested codes $\mathcal{A}_i^M \subset \mathcal{A}_i^{M-1} \subset \cdots \subset \mathcal{A}_i^{i+1} \subset \mathcal{A}_i^i$, where $\mathcal{A}_i^j = [n_i, v_i + \sum_{m=2}^{i-1} \lambda_m^i - \sum_{\ell=i+1}^{j} \lambda_i^\ell, \delta_i^j]_q$ for $i \leqslant j \leqslant M$. Let matrix $H_{\mathcal{A}_i^i}$ represent an $(n_i - v_i - \sum_{m=2}^{i-1} \lambda_m^i) \times n_i$ matrix over $\mathbb{F}_q$ and matrix $H_{\mathcal{A}_i^\ell \setminus \mathcal{A}_i^{\ell-1}}$, $i+1 \leqslant \ell \leqslant j$, represent a $\lambda_i^\ell \times n_i$ matrix over $\mathbb{F}_q$, such that $\mathcal{A}_i^j$ has the following parity-check matrix:

$$H_{\mathcal{A}_i^j} = \begin{bmatrix} H_{\mathcal{A}_i^i} \\ H_{\mathcal{A}_i^{i+1} \setminus \mathcal{A}_i^i} \\ \vdots \\ H_{\mathcal{A}_i^j \setminus \mathcal{A}_i^{j-1}} \end{bmatrix}.$$

For each $2 \leqslant i \leqslant M$, the encoder of code $\mathcal{A}_i^i$ is denoted by $\mathcal{E}_{\mathcal{A}_i^i} : \mathbb{F}_q^{v_i + \sum_{m=2}^{i-1} \lambda_m^i} \to \mathcal{A}_i^i$. We also use $\mathcal{E}_{\mathcal{A}_i^i}^{-1}$ as the inverse of the encoding mapping.

Note that we also define $\mathcal{C}_1^{M+1} = \emptyset$ and $\mathcal{A}_i^{M+1} = \emptyset$ for $2 \leqslant i \leqslant M$.

### A. Construction and Minimum Distance

Now, we give a general algebraic construction for rate-compatible codes $\mathcal{C}_1 \prec \mathcal{C}_2 \prec \cdots \prec \mathcal{C}_M$ by using the nested component codes introduced above.

---

**Construction 1: Encoding Procedure**

**Input:** A length-$k$ vector $\boldsymbol{u}$ of information symbols over $\mathbb{F}_q$.
**Output:** A codeword $\boldsymbol{c}_i \in \mathcal{C}_i$ over $\mathbb{F}_q$, for $i = 1, \ldots, M$.

1: $\boldsymbol{c}_1 = \mathcal{E}_{\mathcal{C}_1}(\boldsymbol{u})$.
2: $\boldsymbol{s}_i = \boldsymbol{c}_1 H_{\mathcal{C}_1^i \setminus \mathcal{C}_1^{i-1}}^T$ for $i = 2, 3, \ldots, M$.
3: **for** $i = 2, \ldots, M$ **do**
4: $\quad \boldsymbol{a}_i^i = \mathcal{E}_{\mathcal{A}_i^i}\left((\boldsymbol{s}_i, \Lambda_2^i, \cdots, \Lambda_{i-1}^i)\right)$. // comment [1] //
5: $\quad \boldsymbol{c}_i = (\boldsymbol{c}_1, \boldsymbol{a}_2^i, \cdots, \boldsymbol{a}_i^i)$.
6: $\quad$ **for** $j = i+1, \ldots, M$ **do**
7: $\quad\quad \Lambda_i^j = \boldsymbol{a}_i^i H_{\mathcal{A}_i^j \setminus \mathcal{A}_i^{j-1}}^T$.
8: $\quad$ **end for**
9: **end for**

---

**Remark 2.** To make Construction 1 clear, consider the case of $M = 3$ as an example. Then a codeword $\boldsymbol{c}_3 \in \mathcal{C}_3$ has the form: $\boldsymbol{c}_3 = \left(\boldsymbol{c}_1, \mathcal{E}_{\mathcal{A}_2^2}(\boldsymbol{s}_2), \mathcal{E}_{\mathcal{A}_3^3}(\boldsymbol{s}_3, \Lambda_2^3)\right)$. The main idea of Construction 1 is to extend the base code $\mathcal{C}_1$ by progressively generating and encoding syndromes of component codes in a proper way. Thus, we call it a *syndrome-coupled* construction.

We have the following theorem on the code parameters of the constructed rate-compatible codes $\mathcal{C}_1 \prec \mathcal{C}_2 \prec \cdots \prec \mathcal{C}_M$.

**Theorem 5.** *From Construction 1, the code $\mathcal{C}_i$, $1 \leqslant i \leqslant M$, has length $N_i = \sum_{j=1}^{i} n_j$ and dimension $K_i = k$. Moreover, assume that $\mathcal{A}_i^j$, $2 \leqslant i \leqslant M$ and $i \leqslant j \leqslant M$, has minimum distance $\delta_i^j \geqslant d_j - d_{i-1}$. Then $\mathcal{C}_i$ has minimum distance*

[1] For $i = 2$, we define $(\boldsymbol{s}_i, \Lambda_2^i, \cdots, \Lambda_{i-1}^i) = \boldsymbol{s}_2$.

$D_i = d_i$. *There exists a decoder for $\mathcal{C}_i$ that can correct any error pattern whose Hamming weight is less than $d_i/2$.*

*Proof:* The code length and dimension are obvious. In the following, we prove the minimum distance. Since the proofs for all $\mathcal{C}_i$, $1 \leqslant i \leqslant M$, are similar, we only give a proof for the code $\mathcal{C}_M$.

We first prove $D_M \geqslant d_M$ by showing that any nonzero codeword $\boldsymbol{c}_M \in \mathcal{C}_M$ has weight at least $d_M$. To see this, for any nonzero codeword $\boldsymbol{c}_1 \in \mathcal{C}_1$, there exists an integer $\gamma_1$, $1 \leqslant \gamma_1 \leqslant M$, such that $\boldsymbol{c}_1 \in \mathcal{C}_1^{\gamma_1}$ and $\boldsymbol{c}_1 \notin \mathcal{C}_1^{\gamma_1+1}$. Let $\boldsymbol{c}_M \in \mathcal{C}_M$ be the codeword derived from $\boldsymbol{c}_1$. Then, we have $w_q(\boldsymbol{c}_M) \geqslant w_q(\boldsymbol{c}_1) \geqslant d_{\gamma_1}$. If $\gamma_1 = M$, we are done; otherwise if $1 \leqslant \gamma_1 \leqslant M-1$ we have $\boldsymbol{s}_{\gamma_1+1} \neq \boldsymbol{0}$ and $\boldsymbol{a}_{\gamma_1+1}^{\gamma_1+1} \neq \boldsymbol{0}$.

Now, for $\boldsymbol{a}_{\gamma_1+1}^{\gamma_1+1}$, there exists an integer $\gamma_2$, $\gamma_1 + 1 \leqslant \gamma_2 \leqslant M$, such that $\boldsymbol{a}_{\gamma_1+1}^{\gamma_1+1} \in \mathcal{A}_{\gamma_1+1}^{\gamma_2}$ and $\boldsymbol{a}_{\gamma_1+1}^{\gamma_1+1} \notin \mathcal{A}_{\gamma_1+1}^{\gamma_2+1}$. Then, we have $w_q(\boldsymbol{c}_M) \geqslant w_q(\boldsymbol{c}_1) + w_q(\boldsymbol{a}_{\gamma_1+1}^{\gamma_1+1}) \geqslant d_{\gamma_1} + d_{\gamma_2} - d_{\gamma_1} = d_{\gamma_2}$. If $\gamma_2 = M$, done; otherwise for $\gamma_1 + 1 \leqslant \gamma_2 \leqslant M-1$, we have $\Lambda_{\gamma_1+1}^{\gamma_2+1} \neq \boldsymbol{0}$ and $\boldsymbol{a}_{\gamma_2+1}^{\gamma_2+1} \neq \boldsymbol{0}$.

Using the same argument as above, it is clear that we can find a sequence of $\gamma_1 < \gamma_2 < \cdots < \gamma_i$, where $i$ is a certain integer $1 \leqslant i \leqslant M$ and $\gamma_i = M$, such that $w_q(\boldsymbol{c}_1) \geqslant d_{\gamma_1}$, $w_q(\boldsymbol{a}_{\gamma_1+1}^{\gamma_1+1}) \geqslant d_{\gamma_2} - d_{\gamma_1}$, $w_q(\boldsymbol{a}_{\gamma_2+1}^{\gamma_2+1}) \geqslant d_{\gamma_3} - d_{\gamma_2}$, $\cdots$, $w_q(\boldsymbol{a}_{\gamma_{i-1}+1}^{\gamma_{i-1}+1}) \geqslant d_{\gamma_i} - d_{\gamma_{i-1}} = d_M - d_{\gamma_{i-1}}$. Then, we have $w_q(\boldsymbol{c}_M) \geqslant w_q(\boldsymbol{c}_1) + \sum_{j=1}^{i-1} w_q(\boldsymbol{a}_{\gamma_j+1}^{\gamma_j+1}) \geqslant d_M$. Thus, we have $D_M \geqslant d_M$.

There exists a codeword $\boldsymbol{c}_1 \in \mathcal{C}_1^M$ such that $w_q(\boldsymbol{c}_1) = d_M$, so we have $w_q(\boldsymbol{c}_M) = d_M$, implying $D_M \leqslant d_M$.

A decoder which can correct any error pattern of Hamming weight less than $d_i/2$ is given in Appendix C. ∎

Next, we provide an example of three-level rate-compatible codes to illustrate Construction 1.

**Example 2.** Consider a set of nested binary BCH codes $\mathcal{C}_1^3 = [15, 5, 7]_2 \subset \mathcal{C}_1^2 = [15, 7, 5]_2 \subset \mathcal{C}_1^1 = [15, 11, 3]_2$. Choose a set of auxiliary codes $\mathcal{A}_2^3 = [5, 1, 4]_2 \subset \mathcal{A}_2^2 = [5, 4, 2]_2$, and $\mathcal{A}_3^3 = [6, 5, 2]_2$, where the code $\mathcal{A}_3^3$ is obtained by shortening an $[8, 4, 4]_2$ extended Hamming code by three information bits.

Then, from Construction 1 and Theorem 5, we obtain three-level rate-compatible codes $\mathcal{C}_1 = [15, 11, 3]_2 \prec \mathcal{C}_2 = [20, 11, 5]_2 \prec \mathcal{C}_3 = [26, 11, 7]_2$. Note that $\mathcal{C}_1$ and $\mathcal{C}_2$ are optimal, achieving the maximum possible dimensions with the given code length and minimum distance. The dimension of $\mathcal{C}_3$ is close to the upper bound 13 according to the online Table [18]. □

### B. Decoding Algorithm and Correctable Error-Erasure Patterns

In the following, we study decoding algorithms and correctable patterns of errors and erasures for rate-compatible codes obtained from Construction 1. For simple notation and concise analysis, we focus on the code $\mathcal{C}_M$. Any results obtained for $\mathcal{C}_M$ can be easily modified for other codes $\mathcal{C}_i$, $1 \leqslant i \leqslant M-1$, so details are omitted.

Assume a codeword $\boldsymbol{c}_M \in \mathcal{C}_M$, $\boldsymbol{c}_M = (\boldsymbol{c}_1, \boldsymbol{a}_2^2, \cdots, \boldsymbol{a}_M^M)$, is transmitted. Let the corresponding received word be $\boldsymbol{y} =$ $(\boldsymbol{y}_1, \boldsymbol{y}_2, \cdots, \boldsymbol{y}_M)$ with errors and erasures, i.e., $\boldsymbol{y} \in (\mathbb{F}_q \cup \{?\})^{N_M}$, where the symbol ? represents an erasure. For $1 \leqslant i \leqslant M$, let $t_i$ and $\tau_i$ denote the number of errors and erasures in the sub-block $\boldsymbol{y}_i$ of the received word $\boldsymbol{y}$.

The code $\mathcal{C}_M$ can correct any combined error and erasure pattern that satisfies the following condition:

$$\begin{aligned} 2t_1 + \tau_1 &\leqslant d_M - 1, \\ 2t_i + \tau_i &\leqslant \delta_i^M - 1, \ \forall \ 2 \leqslant i \leqslant M. \end{aligned} \quad (5)$$

To see this, we present a decoding algorithm, referred to as Algorithm 1, for $\mathcal{C}_M$. It uses the following component error-erasure decoders:

**a)** The error-erasure decoder $\mathcal{D}_{\mathcal{C}_1^i}$ for a coset of the code $\mathcal{C}_1^i$, for $1 \leqslant i \leqslant M$, is defined by

$$\mathcal{D}_{\mathcal{C}_1^i} : (\mathbb{F}_q \cup \{?\})^{n_1} \times (\mathbb{F}_q \cup \{?\})^{\sum_{j=1}^{i} v_j} \to \mathcal{C}_1^i + \boldsymbol{e} \cup \{\text{``e''}\}$$

The decoder $\mathcal{D}_{\mathcal{C}_1^i}$ either produces a codeword in the coset $\mathcal{C}_1^i + \boldsymbol{e}$ or a decoding failure "e". For our purpose, we require that $\mathcal{D}_{\mathcal{C}_1^i}$ have the following error-erasure correcting capability. For a sent codeword $\boldsymbol{c}$ in the coset $\mathcal{C}_1^i + \boldsymbol{e}$, where the vector $\boldsymbol{e}$ is a coset leader, if the inputs of $\mathcal{D}_{\mathcal{C}_1^i}$ are a length-$n_1$ received word $\boldsymbol{y}$ having $t$ errors and $\tau$ erasures, where $2t + \tau \leqslant d_i - 1$, and a correct length-$\sum_{j=1}^{i} v_j$ syndrome vector $\boldsymbol{s}$, $\boldsymbol{s} = \boldsymbol{e} H_{\mathcal{C}_1^i}^T$, then $\mathcal{D}_{\mathcal{C}_1^i}$ can correct all these errors and erasures. It is well known that such a decoder exists [16].

**b)** The error-erasure decoder $\mathcal{D}_{\mathcal{A}_i^j}$ for a coset of the code $\mathcal{A}_i^j$, for $2 \leqslant i \leqslant M$ and $i \leqslant j \leqslant M$, is defined by

$$\mathcal{D}_{\mathcal{A}_i^j} : (\mathbb{F}_q \cup \{?\})^{n_i} \times (\mathbb{F}_q \cup \{?\})^{n_i - v_i - \sum_{m=2}^{i-1} \lambda_m^i + \sum_{\ell=i+1}^{j} \lambda_i^\ell}$$
$$\to \mathcal{A}_i^j + \boldsymbol{e} \cup \{\text{``e''}\}$$

The decoder $\mathcal{D}_{\mathcal{A}_i^j}$ either produces a codeword in the coset $\mathcal{A}_i^j + \boldsymbol{e}$ or a decoding failure "e". Similar to $\mathcal{D}_{\mathcal{C}_1^i}$, we assume that $\mathcal{D}_{\mathcal{A}_i^j}$ has the following error-erasure correcting capability. For a sent codeword $\boldsymbol{c}$ in the coset $\mathcal{A}_i^j + \boldsymbol{e}$, where $\boldsymbol{e}$ is a coset leader, if the inputs of $\mathcal{D}_{\mathcal{A}_i^j}$ are a length-$n_i$ received word $\boldsymbol{y}$ having $t$ errors and $\tau$ erasures, where $2t + \tau \leqslant \delta_i^j - 1$, and a correct length-$(n_i - v_i - \sum_{m=2}^{i-1} \lambda_m^i + \sum_{\ell=i+1}^{j} \lambda_i^\ell)$ syndrome vector $\boldsymbol{s}$, $\boldsymbol{s} = \boldsymbol{e} H_{\mathcal{A}_i^j}^T$, then $\mathcal{D}_{\mathcal{A}_i^j}$ can correct all these errors and erasures.

Now, we present the decoding algorithm as follows.

---

**Algorithm 1: Decoding Procedure for $\mathcal{C}_M$**

**Input:** received word $\boldsymbol{y} = (\boldsymbol{y}_1, \boldsymbol{y}_2, \cdots, \boldsymbol{y}_M)$.
**Output:** A length-$k$ vector $\boldsymbol{u}$ of information symbols over $\mathbb{F}_q$ or a decoding failure "e".

1: **for** $i = M, M-1, \ldots, 2$ **do**
2:   Let the syndrome $\Lambda_i^i = \boldsymbol{0}$.
3:   $\hat{\boldsymbol{a}}_i^i = \mathcal{D}_{\mathcal{A}_i^M}\left(\boldsymbol{y}_i, (\Lambda_i^i, \Lambda_i^{i+1}, \cdots, \Lambda_i^M)\right)$.
4:   $(\boldsymbol{s}_i, \Lambda_2^i, \cdots, \Lambda_{i-1}^i) = \mathcal{E}_{\mathcal{A}_i^i}^{-1}(\hat{\boldsymbol{a}}_i^i)$. // comment [2] //
5: **end for**
6: Let the syndrome $\boldsymbol{s}_1 = \boldsymbol{0}$.

[2]For $i = 2$, we define $(\boldsymbol{s}_i, \Lambda_2^i, \cdots, \Lambda_{i-1}^i) = \boldsymbol{s}_2$.

---

7: $\boldsymbol{c}_1 = \mathcal{D}_{\mathcal{C}_1^M}(\boldsymbol{y}_1, (\boldsymbol{s}_1, \boldsymbol{s}_2, \cdots, \boldsymbol{s}_M))$.
8: Output $\boldsymbol{u} = \mathcal{E}_{\mathcal{C}_1^1}^{-1}(\boldsymbol{c}_1)$ if all above steps are successful; otherwise, return "e".

---

**Theorem 6.** *The code $\mathcal{C}_M$ can correct any combined error and erasure pattern that satisfies the condition in (5), by using Algorithm 1.*

*Proof:* We use Algorithm 1 to decode sub-blocks from $\boldsymbol{y}_M$ to $\boldsymbol{y}_1$. Each sub-block $\boldsymbol{y}_i$ can be corrected successfully due to the condition in (5) and the correcting capability of each component decoder. See Appendix D. ∎

Using nested MDS codes as component codes, Construction 1 can generate an *optimal* code $\mathcal{C}_M$ with respect to the capability of correcting certain error-erasure patterns. For simple notation, we present the case of $M = 3$ as an example.

**Example 3.** Consider a set of nested MDS codes $\mathcal{C}_1^3 = [n_1, n_1 - d_3 + 1, d_3]_q \subset \mathcal{C}_1^2 = [n_1, n_1 - d_2 + 1, d_2]_q \subset \mathcal{C}_1^1 = [n_1, n_1 - d_1 + 1, d_1]_q$. Choose a set of auxiliary MDS codes $\mathcal{A}_2^3 = [2(d_2 - d_1) - 1, 2d_2 - d_3 - d_1, d_3 - d_1]_q \subset \mathcal{A}_2^2 = [2(d_2 - d_1) - 1, d_2 - d_1, d_2 - d_1]_q$, and $\mathcal{A}_3^3 = [3(d_3 - d_2) - 1, 2(d_3 - d_2), d_3 - d_2]_q$.

Then, from Construction 1 and Theorem 5, we obtain three-level rate-compatible codes $\mathcal{C}_1 = [n_1, n_1 - d_1 + 1, d_1]_q \prec \mathcal{C}_2 = [n_1 + 2(d_2 - d_1) - 1, n_1 - d_1 + 1, d_2]_q \prec \mathcal{C}_3 = [n_1 + 2(d_2 - d_1) + 3(d_3 - d_2) - 2, n_1 - d_1 + 1, d_3]_q$.

From the condition in (5) and Theorem 6, the code $\mathcal{C}_3$ can correct any pattern of errors and erasures satisfying

$$2t_i + \tau_i \leqslant d_3 - d_{i-1} - 1, \ \forall \ 1 \leqslant i \leqslant 3, \quad (6)$$

where $d_0$ is defined to be 0.

In general, the dimension of $\mathcal{C}_3$ cannot achieve the upper bounds given by traditional bounds (e.g., Singleton and Hamming bounds). However, $\mathcal{C}_3$ is optimal in the sense of having the largest possible dimension among all codes with the three-level structure and the same error-erasure correcting capability; that is, we have the following lemma, whose proof is in Appendix E.

**Lemma 7.** *Let $\mathcal{C}_3$ be a code of length $n_1 + 2(d_2 - d_1) + 3(d_3 - d_2) - 2$ and dimension $k_3$ over $\mathbb{F}_q$. Each codeword $\boldsymbol{c}_3 \in \mathcal{C}_3$ has three sub-blocks $(\boldsymbol{c}_1, \boldsymbol{a}_2^2, \boldsymbol{a}_3^3)$: 1) $\boldsymbol{c}_1$ of length $n_1$, 2) $\boldsymbol{a}_2^2$ of length $2(d_2 - d_1) - 1$, and 3) $\boldsymbol{a}_3^3$ of length $3(d_3 - d_2) - 1$. Assume that each sub-block of $\mathcal{C}_3$ can correct all error and erasure patterns satisfying the condition in (6). Then, we must have $k_3 \leqslant n_1 - d_1 + 1$.* □

In Algorithm 1, the code $\mathcal{C}_M$ is decoded by $M$ steps, so we can bound the decoding error probability $P_e^{(N_M)}(\mathcal{C}_M)$ of $\mathcal{C}_M$ by the decoding error probability of each step as

$$P_e^{(N_M)}(\mathcal{C}_M) \leqslant 1 - \left(1 - P_e^{(n_1)}(\mathcal{C}_1^M)\right) \prod_{i=2}^{M} \left(1 - P_e^{(n_i)}(\mathcal{A}_i^M)\right),$$

which provides a fast way to predict the performance of $\mathcal{C}_M$. In particular, if each component code is (shortened) BCH code, then $P_e^{(N_M)}(\mathcal{C}_M)$ can be easily estimated by some calculations. We use a simple example to illustrate this estimation.

**Example 4.** Consider two nested binary BCH codes $\mathcal{C}_1^2 = [8191, 7411]_2 \subset \mathcal{C}_1^1 = [8191, 7671]_2$. The codes $\mathcal{C}_1^1$ and $\mathcal{C}_1^2$ can correct 40 and 60 errors, respectively. Choose an auxiliary shortened BCH code $\mathcal{A}_2^2 = [359, 260]_2$, which can correct 11 errors. Then, from Construction 1, we obtain two-level rate-compatible codes $\mathcal{C}_1 = [8191, 7671]_2 \prec \mathcal{C}_2 = [8550, 7671]_2$. Now, send $\mathcal{C}_2$ over a binary symmetric channel (BSC) with crossover probability $p$. The error probability of $\mathcal{C}_2$ satisfies

$$P_e^{(N_2)}(\mathcal{C}_2) \leqslant 1 - \left(1 - P_e^{(n_1)}(\mathcal{C}_1^2)\right)\left(1 - P_e^{(n_2)}(\mathcal{A}_2^2)\right)$$

$$\leqslant 1 - \left(\sum_{i=0}^{t_1} \binom{n_1}{i} p^i (1-p)^{n_1-i}\right)\left(\sum_{i=0}^{t_2} \binom{n_2}{i} p^i (1-p)^{n_2-i}\right),$$

where $N_2 = 8550$, $n_1 = 8191$, $n_2 = 359$, $t_1 = 60$, and $t_2 = 11$. For instance, for $p = 0.0035$, we compute $P_e^{(N_2)}(\mathcal{C}_2) \leqslant 1.049 \times 10^{-7}$; for $p = 0.004$, we have $P_e^{(N_2)}(\mathcal{C}_2) \leqslant 6.374 \times 10^{-6}$. For $p \geqslant 0.0035$, the performance of $\mathcal{C}_2$ (rate 0.8972) is comparable to, although still worse than, a shortened $[8553, 7671]_2$ BCH code $\mathcal{C}_2'$ that has rate 0.8969 and can correct 63 errors. For instance, for $p = 0.0035$ and 0.004, $\mathcal{C}_2'$ has error probabilities $4.035 \times 10^{-8}$ and $3.315 \times 10^{-6}$. □

## V. Capacity-Achieving Rate-Compatible Codes

In this section, we show that if we choose component codes properly, Construction 1 can generate rate-compatible codes which achieve the capacities of a set of degraded $q$-ary symmetric channels simultaneously.

More specifically, consider a set of $M$ degraded $q$-ary symmetric channels $W_1 \succ W_2 \succ \cdots \succ W_M$ with crossover probabilities $p_1 < p_2 < \cdots < p_M$ respectively, where $p_1 > 0$ and $p_M < 1 - (1/q)$. Let $C(W_i)$ denote the capacity of the channel $W_i$, i.e., $C(W_i) = 1 - H_q(p_i)$. It is clear that $C(W_1) > C(W_2) > \cdots > C(W_M)$. For any rates $R_1 > R_2 > \cdots > R_M$ such that $R_i < C(W_i)$ for all $1 \leqslant i \leqslant M$, we will show that Construction 1 can generate rate-compatible codes $\mathcal{C}_1 \prec \mathcal{C}_2 \prec \cdots \prec \mathcal{C}_M$ where $\mathcal{C}_i = [N_i, R_i N_i]_q$ such that the decoding error probability of each $\mathcal{C}_i$ over channel $W_i$ satisfies $P_e^{(N_i)}(\mathcal{C}_i) \to 0$, as $N_i$ goes to infinity.

To this end, we first present the following lemma on the existence of nested capacity-achieving linear codes. Its proof can be found in Appendix F.

**Lemma 8.** *Consider a set of $M$ degraded $q$-ary symmetric channels $W_1 \succ W_2 \succ \cdots \succ W_M$ with capacities $C(W_1) > C(W_2) > \cdots > C(W_M)$. For any rates $R_1 > R_2 > \cdots > R_M$ such that $R_i < C(W_i)$, there exists a sequence of nested linear codes $\mathcal{C}_1^M = [n, k_M = R_M n]_q \subset \mathcal{C}_1^{M-1} = [n, k_{M-1} = R_{M-1} n]_q \subset \cdots \subset \mathcal{C}_1^1 = [n, k_1 = R_1 n]_q$ such that the decoding error probability of each $\mathcal{C}_1^i$ over channel $W_i$, under nearest-codeword (ML) decoding, satisfies $P_e^{(n)}(\mathcal{C}_1^i) \to 0$, as $n$ goes to infinity.*

Now, we are ready to construct capacity-achieving rate-compatible codes from Construction 1. To do so, we choose a set of nested capacity-achieving codes to be the component codes, which exist according to Lemma 8.

**1)** Choose a set of nested capacity-achieving codes $\mathcal{C}_1^M \subset \mathcal{C}_1^{M-1} \subset \cdots \subset \mathcal{C}_1^1 = \mathcal{C}_1 = [n_1, k]_q$, where $\mathcal{C}_1^i = [n_1, n_1 - \sum_{m=1}^{i} v_m]_q$ for $1 \leqslant i \leqslant M$. Let $\mathcal{C}_1^i$ have the required rate $R_i < C(W_i)$, and for $\mathcal{C}_1^i$ over channel $W_i$, its error probability satisfies $P_e^{(n_1)}(\mathcal{C}_1^i) \to 0$, as $n_1$ goes to infinity.

**2)** For $i$th level, $2 \leqslant i \leqslant M$, choose a set of auxiliary nested capacity-achieving codes $\mathcal{A}_i^M \subset \mathcal{A}_i^{M-1} \subset \cdots \subset \mathcal{A}_i^{i+1} \subset \mathcal{A}_i^i$, where $\mathcal{A}_i^j = [n_i, v_i + \sum_{m=2}^{i-1} \lambda_m^i - \sum_{\ell=i+1}^{j} \lambda_i^\ell]_q$ for $i \leqslant j \leqslant M$. Let $\mathcal{A}_i^j$ have the required rate $R_j < C(W_j)$, and for $\mathcal{A}_i^j$ over channel $W_j$, the decoding error probability satisfies $P_e^{(n_i)}(\mathcal{A}_i^j) \to 0$, as $n_i$ goes to infinity.

Note that compared to Section IV, here we care about rate and capacity-achieving property, instead of minimum distance, of each component code.

**Theorem 9.** *With the above component codes, from Construction 1, we obtain a sequence of rate-compatible codes $\mathcal{C}_1 \prec \mathcal{C}_2 \prec \cdots \prec \mathcal{C}_M$, where $\mathcal{C}_i$, $1 \leqslant i \leqslant M$, has length $N_i = \sum_{j=1}^{i} n_j$, dimension $K_i = k$, and rate $R_i$. Moreover, for each $\mathcal{C}_i$ over channel $W_i$, it is capacity-achieving, i.e., the error probability $P_e^{(N_i)}(\mathcal{C}_i) \to 0$, as $N_i$ goes to infinity.*

*Proof:* The proof has two parts. First, we need to prove the rate of $\mathcal{C}_i$ is $R_i$. Second, we will show that the code $\mathcal{C}_i$ can be decoded by $i$ steps. For each step, the decoding error probability goes to zero, as the code length of $\mathcal{C}_i$ goes to infinity. Thus, the error probability $P_e^{(N_i)}(\mathcal{C}_i) \to 0$, as $N_i$ goes to infinity. See Appendix G. ■

**Remark 3.** Polar codes were proved to have the nested capacity-achieving property [9]. Thus, they can be used as the component codes to construct capacity-achieving rate-compatible codes.

When we were preparing this work, we found recent independent works on capacity-achieving rateless and rate-compatible codes based on polar codes [11] [8]. By investigating the construction in [8] carefully, we find our construction with polar codes as component codes is equivalent to theirs by one-to-one mapping the *syndrome* in our construction to the *frozen bits* in their construction by a *full rank* matrix; see Appendix H for the proof. Since the construction in [8] is based on the generator matrix, our construction can be seen as another interpretation of their construction from a parity-check matrix perspective.

## VI. Conclusion

This work proposed a new algebraic construction for generating rate-compatible codes with increasing minimum distances. We also proved that our construction can be capacity-achieving by using proper component codes, validating the optimality of the construction. With polar codes as component codes, the equivalence between our construction and the one in [8] was identified.

Our construction is very general. Many linear codes (e.g., BCH, RS, and LDPC codes) can be used as its component codes, and some of them were shown as examples. Our parity-check matrix based approach enables us to conveniently obtain the combinatorial property (e.g., minimum distance) of the constructed rate-compatible codes, as well as their decoders.

## Acknowledgment

This work was supported by Seagate Technology and NSF Grants CCF-1405119 and CCF-1619053.

## APPENDIX A
## PROOF OF THEOREM 2

*Proof:* We first define an $(n_M - k) \times n_M$ matrix $\Phi_M$ over $\mathbb{F}_q$ in the following block lower triangular form,

$$\Phi_M = \begin{bmatrix} H_{1,1} & \mathbf{0} & \ldots & \mathbf{0} & \mathbf{0} \\ H_{2,1} & H_{2,2} & \ldots & \mathbf{0} & \mathbf{0} \\ \vdots & \vdots & \ddots & \vdots & \vdots \\ H_{M-1,1} & H_{M-1,2} & \ldots & H_{M-1,M-1} & \mathbf{0} \\ H_{M,1} & H_{M,2} & \ldots & H_{M,M-1} & H_{M,M} \end{bmatrix}, \quad (7)$$

where $H_{1,1}$ is an $(n_1 - k) \times n_1$ matrix. For $2 \leqslant i \leqslant M$, matrix $H_{i,1}$ has size $r_i \times n_1$. For $2 \leqslant i \leqslant M$ and $2 \leqslant j \leqslant i$, matrix $H_{i,j}$ has size $r_i \times r_j$.

For $1 \leqslant i \leqslant M$, we assign the upper left $(n_i - k) \times n_i$ submatrix of $\Phi_M$, denoted by $\Phi_i$, to be the parity-check matrix of $C_i$. For example, matrices $H_{1,1}$ and $\Phi_M$ are parity-check matrices of $C_1$ and $C_M$, respectively.

Now, we show how to construct $H_{i,j}$, $1 \leqslant i \leqslant M$ and $1 \leqslant j \leqslant i$, such that each code $C_i$ has its desired code parameters.

First, for $2 \leqslant i \leqslant M$, we choose $H_{i,i}$ to be an $r_i \times r_i$ identity matrix. For $3 \leqslant i \leqslant M$ and $2 \leqslant j \leqslant i-1$, we choose matrix $H_{i,j}$ to be an arbitrary matrix in $\mathbb{F}_q^{r_i \times r_j}$. Next, we construct columns of $H_{i,1}$, $1 \leqslant i \leqslant M$, iteratively, as the technique used in the proof of the GV bound. We use $\boldsymbol{h}_\ell(i)$, $1 \leqslant \ell \leqslant n_1$ and $1 \leqslant i \leqslant M$, to denote the $\ell$th column of matrix $\Phi_i$ which is the parity-check matrix of $C_i$. Assume that we have already added the leftmost $\ell - 1$ columns of matrix $\Phi_M$. In order to show that in $\mathbb{F}_q^{n_M - k}$ there is a vector that can be used as the $\ell$th column $\boldsymbol{h}_\ell(M)$ of matrix $\Phi_M$, we only need to show that the total number of bad vectors is less than $q^{n_M - k}$. We count the number of bad vectors as follows.

For code $C_1$, it requires that every $d_1 - 1$ columns in $\Phi_1$ are linearly independent. A bad vector for the $\ell$th column $\boldsymbol{h}_\ell(1)$ in $\Phi_1$ is a vector that can be expressed as a linear combination of $d_1 - 2$ columns in the preceding $\ell - 1$ columns. There are a total of $\sum_{m=0}^{d_1 - 2} \binom{\ell-1}{m}(q-1)^m$ such bad vectors, so we exclude $N_1(\ell) = \sum_{m=0}^{d_1 - 2} \binom{\ell-1}{m}(q-1)^m \times q^{\sum_{j=2}^M r_j}$ bad vectors for the column $\boldsymbol{h}_\ell(M)$.

Similarly, for code $C_i$, $2 \leqslant i \leqslant M$, it requires that every $d_i - 1$ columns in $\Phi_i$ are linearly independent. A bad vector for the $\ell$th column $\boldsymbol{h}_\ell(i)$ in $\Phi_i$ is a vector that can be expressed as a linear combination of $d_i - 2$ columns in the preceding $\ell - 1 + \sum_{j=2}^i r_j$ selected columns, so we have a total of $\sum_{m=0}^{d_i - 2} \binom{\ell - 1 + \sum_{j=2}^i r_j}{m}(q-1)^m$ such bad vectors. Then, we exclude $N_i(\ell) = \sum_{m=0}^{d_i - 2} \binom{\ell - 1 + \sum_{j=2}^i r_j}{m}(q-1)^m \times q^{\sum_{x=i+1}^M r_x}$ bad vectors for the column $\boldsymbol{h}_\ell(M)$.

Since we assume that the inequalities (2) are satisfied, we have $N_i(\ell) < \frac{q^{n_M - k}}{M}$ for $1 \leqslant i \leqslant M$ and $1 \leqslant \ell \leqslant n_1$. Thus, we have $\sum_{i=1}^M N_i(\ell) < q^{n_M - k}$, which indicates that a good column $\boldsymbol{h}_\ell(M)$ can be found. ∎

## APPENDIX B
## PROOF OF THEOREM 4

*Proof:* Let an $(n_1 - k) \times n_1$ matrix $H_1$ represent the parity-check matrix of $C_1$. Assume that $C_2$ has a parity-check matrix $H_2$ in the form

$$H_2 = \begin{bmatrix} H_1 & \mathbf{0} \\ H & I \end{bmatrix}, \quad (8)$$

where $H$ is an $r_2 \times n_1$ matrix and matrix $I$ represents an $r_2 \times r_2$ identity matrix. Construct an ensemble of $(n_2 - k) \times n_2$ matrices $\{H_2\}$ by using all $r_2 \times n_1$ matrices $H$ over $\mathbb{F}_q$. We then assume a uniform distribution over the ensemble $\{H_2\}$.

We say a matrix $H_2$ is bad, if there exists a vector $\boldsymbol{x} \in \mathbb{F}_q^{n_2}$ such that $\boldsymbol{x} H_2^T = \mathbf{0}$ and $0 < w_q(\boldsymbol{x}) < d_2$. Thus, we only need to prove the probability $\Pr\{H_2 \text{ is bad}\} < 1$, i.e., not all $H_2$ are bad.

Define sets $\mathcal{B}' = \{\boldsymbol{x} \in \mathbb{F}_q^{n_2} : \boldsymbol{x}[H_1, \mathbf{0}]^T = \mathbf{0}\}$, $\mathcal{B}'' = \{\boldsymbol{x} \in \mathcal{B}' : w_q(\boldsymbol{x}) > 0,$ and the leading nonzero entry of $\boldsymbol{x}$ is $1\}$, and $\mathcal{B} = \{\boldsymbol{x} \in \mathcal{B}'' : w_q(\pi_{[n_1]}(\boldsymbol{x})) > 0\}$. We also define $B_w = |\{\boldsymbol{x} \in \mathcal{B} : w_q(\boldsymbol{x}) = w\}|$. It is clear that $B_w = \frac{1}{q-1} \sum_{m=1}^w A_m \binom{r_2}{w-m}(q-1)^{w-m}$, for $1 \leqslant w \leqslant n_2$.

Now, we have

$$\Pr\{H_2 \text{ is bad}\}$$
$$= \Pr\{\text{For some } x \in \mathcal{B}', \ 0 < w_q(x) < d_2, \ x[H, I]^T = \mathbf{0}\}$$
$$= \Pr\{\text{For some } x \in \mathcal{B}'', \ 0 < w_q(x) < d_2, \ x[H, I]^T = \mathbf{0}\}$$
$$= \Pr\{\text{For some } x \in \mathcal{B}, \ 0 < w_q(x) < d_2, \ x[H, I]^T = \mathbf{0}\}$$
$$\overset{(a)}{\leqslant} \sum_{x \in \mathcal{B} \text{ and } 0 < w_q(x) < d_2} \Pr\{x[H, I]^T = \mathbf{0}\}$$
$$= \frac{\sum_{w=1}^{d_2-1} B_w}{q^{r_2}},$$

where step (a) follows from the union bound. ∎

## APPENDIX C
## PROOF FOR PART OF THEOREM 5

*Proof:* For notational simplicity, we give a decoder for the code $\mathcal{C}_M$ that can correct any error pattern whose Hamming weight is less than $d_M/2$. The decoder for $\mathcal{C}_M$ can be easily modified for other codes $\mathcal{C}_i$, $1 \leqslant i \leqslant M-1$, correspondingly, so are omitted.

We present an error decoding algorithm, referred to as Algorithm 2, for $\mathcal{C}_M$. It uses the following nearest-codeword decoders:

**a)** The nearest-codeword decoder $\mathscr{D}_{\mathcal{C}_1^i}$ for a coset of the code $\mathcal{C}_1^i$, for $1 \leqslant i \leqslant M$, is defined by

$$\mathscr{D}_{\mathcal{C}_1^i} : \mathbb{F}_q^{n_1} \times \mathbb{F}_q^{\sum_{j=1}^i v_j} \to \mathcal{C}_1^i + e$$

according to the following decoding rules: for a length-$n_1$ input vector $y$, and a length-$\sum_{j=1}^i v_j$ syndrome vector $s$, if $c$ is a closest codeword to $y$ in the coset $\mathcal{C}_1^i + e$, where the vector $e$ is a coset leader determined by both the code $\mathcal{C}_1^i$ and the syndrome vector $s$, i.e., $s = eH_{\mathcal{C}_1^i}^T$, then $\mathscr{D}_{\mathcal{C}_1^i}(y, s) = c$.

**b)** The nearest-codeword decoder $\mathscr{D}_{\mathcal{A}_i^j}$ for a coset of the code $\mathcal{A}_i^j$, for $2 \leqslant i \leqslant M$ and $i \leqslant j \leqslant M$, is defined by

$$\mathscr{D}_{\mathcal{A}_i^j} : \mathbb{F}_q^{n_i} \times \mathbb{F}_q^{n_i - v_i - \sum_{m=2}^{i-1} \lambda_m^i + \sum_{\ell=i+1}^{j} \lambda_i^\ell} \to \mathcal{A}_i^j + e$$

according to the following decoding rules: for a length-$n_i$ input vector $y$, and a length-$(n_i - v_i - \sum_{m=2}^{i-1} \lambda_m^i + \sum_{\ell=i+1}^{j} \lambda_i^\ell)$ syndrome vector $s$, if $c$ is a closest codeword to $y$ in the coset $\mathcal{A}_i^j + e$, where the vector $e$ is a coset leader determined by both the code $\mathcal{A}_i^j$ and the syndrome vector $s$, i.e., $s = eH_{\mathcal{A}_i^j}^T$, then $\mathscr{D}_{\mathcal{A}_i^j}(y, s) = c$.

The input to Algorithm 2 is a received word $y = (y_1, y_2, \cdots, y_M)$, $y \in \mathbb{F}_q^{N_M}$, corresponding to a transmitted codeword $c_M \in \mathcal{C}_M$, i.e.,

$$c_M = (c_1, a_2^2, \cdots, a_M^M).$$

Assume that the Hamming distance between $y$ and $c_M$ satisfies $d_q(y, c_M) < d_M/2$. Then Algorithm 2 in the following will output the correct codeword $c_M$.

---

**Algorithm 2: Decoding Procedure**

**Input:** received word $y = (y_1, y_2, \cdots, y_M)$.

**Output:** codeword $c_M \in \mathcal{C}_M$ or a decoding failure "e".

**Level 1:**
1: Let the syndrome $\hat{s}_1 = \mathbf{0}$ and $\hat{c}_1 = \mathscr{D}_{\mathcal{C}_1^1}(y_1, \hat{s}_1)$.
2: check$(\hat{c}_1, y)$.

**Level 2 − Level $M$:**
1: **for** $i = 2, 3, \cdots, M$ **do**
2:   **for** $j = i, i-1, \ldots, 2$ **do**
3:     Let the syndrome $\hat{\Lambda}_j^j = \mathbf{0}$.
4:     $\hat{a}_j^j = \mathscr{D}_{\mathcal{A}_j^i}\left(y_j, (\hat{\Lambda}_j^j, \hat{\Lambda}_j^{j+1}, \cdots, \hat{\Lambda}_j^i)\right)$.
5:     $(\hat{s}_j, \hat{\Lambda}_2^j, \cdots, \hat{\Lambda}_{j-1}^j) = \mathcal{E}_{\mathcal{A}_j^j}^{-1}(\hat{a}_j^j)$.   // comment [3] //
6:   **end for**
    $\hat{c}_1 = \mathscr{D}_{\mathcal{C}_1^i}\left(y_1, (\hat{s}_1, \hat{s}_2, \cdots, \hat{s}_i)\right)$.
7:   check$(\hat{c}_1, y)$.
8: **end for**
9: If no codeword $c_M$ has been produced in the above steps, then return "e".

---

The check function in Algorithm 2 is defined as follows.

---

**The check function**

**Input:** $\hat{c}_1$ and received word $y$.

1: Let $\hat{u} = \mathcal{E}_{\mathcal{C}_1^1}^{-1}(\hat{c}_1)$, and encode $\hat{u}$ using Construction 1 to obtain $c_M \in \mathcal{C}_M$.
2: If $d_q(y, c_M) < d_M/2$, then output $c_M$ and exit the Decoding Procedure.

---

We use check function to check whether the Hamming distance between the computed codeword $c_M$ and the received word $y$ is less than $d_M/2$. Since the hypothesis is that the number of errors is less than $d_M/2$, only the transmitted codeword will pass this check.

Now, we prove that Algorithm 2 can correct any error pattern whose Hamming weight is less than $d_M/2$. Let $T(v)$ denote the number of errors occurred in the vector $v$. We use induction for the proof.

We propose the following *claim*: if the decoder in Algorithm 2 proceeds at the $j$th level and a total number of errors among $y_1, y_2, \cdots, y_j$ is less than $\frac{d_j}{2}$, it can decode all these errors, producing the correct codeword $c_M$ successfully, and stops proceeding. We only need to prove this claim is true for all $1 \leqslant j \leqslant M$.

For $j = 1$, if $T(y_1) < \frac{d_1}{2}$, then in Level 1 the correct codeword $c_M$ will be produced, so the claim holds for $j = 1$.

For $j = 2$, since the decoder proceeds at this level, it indicates that $T(y_1) \geqslant \frac{d_1}{2}$. If $T(y_1, y_2) < \frac{d_2}{2}$, then $T(y_2) < \frac{d_2 - d_1}{2}$ due to $T(y_1) \geqslant \frac{d_1}{2}$. Then the correct syndrome $s_2$ can be obtained from $y_2$, and then $y_1$ can be decoded correctly into $c_1$ by using $s_2$. Thus, the correct codeword $c_M$ will be produced, so the claim holds for $j = 2$.

Now, we prove that if the claim is true for $1, 2, \cdots, j-1$, then it is also true for $j$. If the decoder proceeds at $j$th level, then it means that $T(y_1, y_2, \cdots, y_{j-1}) \geqslant \frac{d_{j-1}}{2}$.

---

[3] For $j = 2$, we define $(\hat{s}_j, \hat{\Lambda}_2^j, \cdots, \hat{\Lambda}_{j-1}^j) = \hat{s}_2$.

Since we assume that $T(\boldsymbol{y}_1, \boldsymbol{y}_2, \cdots, \boldsymbol{y}_{j-1}, \boldsymbol{y}_j) < \frac{d_j}{2}$, we have $T(\boldsymbol{y}_j) < \frac{d_j - d_{j-1}}{2}$. Then, the correct syndromes $\boldsymbol{s}_j, \Lambda_2^j, \cdots, \Lambda_{j-1}^j$ will be obtained by decoding $\boldsymbol{y}_j$ correctly. Since the decoder proceeds at the $j$th level, it also means that $T(\boldsymbol{y}_1, \boldsymbol{y}_2, \cdots, \boldsymbol{y}_{j-2}) \geqslant \frac{d_{j-2}}{2}$, so we have $T(\boldsymbol{y}_{j-1}) < \frac{d_j - d_{j-2}}{2}$. Since we already have $\Lambda_{j-1}^j$, then $\boldsymbol{y}_{j-1}$ can be decoded correctly.

Similarly, $T(\boldsymbol{y}_\ell) < \frac{d_j - d_{\ell-1}}{2}$ for $\ell = j-2, \cdots, 2$, so $\boldsymbol{y}_\ell$ can be decoded correctly, using $\Lambda_\ell^{\ell+1}, \Lambda_\ell^{\ell+2}, \cdots, \Lambda_\ell^j$. Now, we have obtained all correct syndromes $\boldsymbol{s}_2, \boldsymbol{s}_3, \cdots, \boldsymbol{s}_j$. Since $T(\boldsymbol{y}_1) < \frac{d_j}{2}$, then $\boldsymbol{y}_1$ can be decoded successfully with all these correct syndromes. Thus, the correct codeword $\boldsymbol{c}_M$ will be produced, and we complete the proof. ■

## APPENDIX D
## PROOF OF THEOREM 6

*Proof:* The proof follows from Algorithm 1 that decodes the last sub-block $\boldsymbol{y}_M$ to the first sub-block $\boldsymbol{y}_1$ progressively. First, since the code $\mathcal{A}_M^M$ has minimum distance $\delta_M^M$, it can correct $\boldsymbol{y}_M$ under the condition $2t_M + \tau_M \leqslant \delta_M^M - 1$. Thus, we obtain correct syndromes $\boldsymbol{s}_M, \Lambda_2^M, \cdots, \Lambda_{M-1}^M$.

Next, with the correct syndrome $\Lambda_{M-1}^M$, the coset decoder $\mathcal{D}_{\mathcal{A}_{M-1}^M}$ can correct $\boldsymbol{y}_{M-1}$ under the condition $2t_{M-1} + \tau_{M-1} \leqslant \delta_{M-1}^M - 1$. Thus, we obtain correct syndromes $\boldsymbol{s}_{M-1}, \Lambda_2^{M-1}, \cdots, \Lambda_{M-2}^{M-1}$.

Conduct above decoding procedure progressively. For any $i$, $2 \leqslant i \leqslant M-2$, using the correct syndromes $\Lambda_i^{i+1}, \cdots, \Lambda_i^M$ for coset decoding, the sub-block $\boldsymbol{y}_i$ can be corrected under the condition $2t_i + \tau_i \leqslant \delta_i^M - 1$.

At the last step, we have obtained correct syndromes $\boldsymbol{s}_2, \cdots, \boldsymbol{s}_M$. Therefore, the sub-block $\boldsymbol{y}_1$ is corrected. ■

## APPENDIX E
## PROOF OF LEMMA 7

*Proof:* We prove Lemma 7 by contradiction.

Let $\mathcal{I}_1$ be the set of any $d_3 - 1$ coordinates of $\boldsymbol{c}_1$, $\mathcal{I}_2$ be the set of any $d_3 - d_1 - 1$ coordinates of $\boldsymbol{a}_2^2$, and $\mathcal{I}_3$ be the set of any $d_3 - d_2 - 1$ coordinates of $\boldsymbol{a}_3^3$. Let $\mathcal{I}$ be the set of all the coordinates of $\boldsymbol{c}_3$.

We have $|\mathcal{I} \setminus (\mathcal{I}_1 \cup \mathcal{I}_2 \cup \mathcal{I}_3)| = n_1 - d_1 + 1$. Now, assume that $k_3 > n_1 - d_1 + 1$. Then, there exist at least two distinct codewords $\boldsymbol{c}_3'$ and $\boldsymbol{c}_3''$ in $\mathcal{C}_3$ that agree on the coordinates in the set $\{i : i \in \mathcal{I} \setminus (\mathcal{I}_1 \cup \mathcal{I}_2 \cup \mathcal{I}_3)\}$. We erase the values on the coordinates in $\{i : i \in \mathcal{I}_1 \cup \mathcal{I}_2 \cup \mathcal{I}_3\}$ of both $\boldsymbol{c}_3'$ and $\boldsymbol{c}_3''$. This erasure pattern satisfies the condition in (6). Since $\boldsymbol{c}_3'$ and $\boldsymbol{c}_3''$ are distinct, this erasure pattern is uncorrectable. Thus, our assumption that $k_3 > n_1 - d_1 + 1$ is violated. ■

## APPENDIX F
## PROOF OF LEMMA 8

*Proof:* To prove the lemma, we need two known results from [16]. We state them as follows.

**Theorem 10.** *For the $q$-ary symmetric channel with crossover probability $p$, $p \in (0, 1 - (1/q))$, let $n$ and $nR$ be integers such that $R < 1 - H_q(p)$. Let $\overline{P_e^{(n)}(\mathcal{C})}$ denote the average of $P_e^{(n)}(\mathcal{C})$ over all linear $[n, nR]_q$ codes $\mathcal{C}$ with nearest-codeword decoding. Then,*

$$\overline{P_e^{(n)}(\mathcal{C})} < 2q^{-nE_q(p,R)},$$

*where $E_q(p, R) > 0$.*

The Theorem 10 gives the following theorem.

**Theorem 11.** *For every $\rho \in (0, 1]$, all but a fraction less than $\rho$ of the linear $[n, nR]_q$ codes $\mathcal{C}$ satisfy*

$$P_e^{(n)}(\mathcal{C}) < (1/\rho) 2q^{-nE_q(p,R)}.$$

Now, we are ready to prove Lemma 8. Consider an ensemble $\mathcal{G}_1$ of all $k_1 \times n$ full rank matrices over $\mathbb{F}_q$. The size of $\mathcal{G}_1$ is $|\mathcal{G}_1| = (q^n - 1)(q^n - q) \cdots (q^n - q^{k_1 - 1})$. Now, for each matrix $G_1^i \in \mathcal{G}_1$, $1 \leqslant i \leqslant |\mathcal{G}_1|$, take the lowest $k_2$ rows to form a new matrix $G_2^i$. All these new matrices form a new ensemble $\mathcal{G}_2$. It is clear that $|\mathcal{G}_2| = |\mathcal{G}_1|$ and in $\mathcal{G}_2$, each $k_2 \times n$ full rank matrix over $\mathbb{F}_q$ has $(q^n - q^{k_2})(q^n - q^{k_2 + 1}) \cdots (q^n - q^{k_1 - 1})$ copies. Similarly, for each matrix $G_1^i \in \mathcal{G}_1$, $1 \leqslant i \leqslant |\mathcal{G}_1|$, take the lowest $k_j$, $3 \leqslant j \leqslant M$, rows to form a new matrix $G_j^i$. All these new matrices form a new ensemble $\mathcal{G}_j$. It is clear that $|\mathcal{G}_j| = |\mathcal{G}_1|$ and in $\mathcal{G}_j$, each $k_j \times n$ full rank matrix over $\mathbb{F}_q$ has $(q^n - q^{k_j})(q^n - q^{k_j + 1}) \cdots (q^n - q^{k_1 - 1})$ copies.

Note that each linear $[n, k]_q$ code has the same number of generator matrices. Therefore, from Theorem 11, in each ensemble $\mathcal{G}_j$ for $1 \leqslant j \leqslant M$, we have at least $x$ fraction of all matrices in this ensemble will generate linear codes $\mathcal{C}$ such that the error probability $P_e^{(n)}(\mathcal{C}) < (\frac{1}{1-x}) 2q^{-nE_q(p_j, R_j)}$.

Now, it is not hard to see that in $\mathcal{G}_1$ we can find a subset $\overline{\mathcal{G}}_1 \subseteq \mathcal{G}_1$ such that $\overline{\mathcal{G}}_1$ has at least $Mx - (M - 1)$ fraction of all matrices in $\mathcal{G}_1$, and for each matrix $\overline{G}_1$ in $\overline{\mathcal{G}}_1$, for all $1 \leqslant j \leqslant M$, the lowest $k_j$ rows of $\overline{G}_1$ will generate linear codes $\mathcal{C}_1^j$ with the error probability $P_e^{(n)}(\mathcal{C}_1^j) < (\frac{1}{1-x}) 2q^{-nE_q(p_j, R_j)}$.

Choosing any $x$ satisfying $\frac{M-1}{M} < x < 1$, it is clear that there exists a sequence of nested linear codes $\mathcal{C}_1^M = [n, k_M = R_M n]_q \subset \mathcal{C}_1^{M-1} = [n, k_{M-1} = R_{M-1} n]_q \subset \cdots \subset \mathcal{C}_1^1 = [n, k_1 = R_1 n]_q$ such that for all $1 \leqslant i \leqslant M$, the error probability $P_e^{(n)}(\mathcal{C}_1^i) \to 0$, as $n$ goes to infinity. ■

## APPENDIX G
## PROOF OF THEOREM 9

*Proof:* The code length and dimension are obvious. In the following, we prove the rate; that is, to show $\frac{k}{N_i} = \frac{k}{\sum_{j=1}^i n_j} = R_i$. For $i = 1$, it is trivial, since the rate of $\mathcal{C}_1^1$ is $R_1$. For $i = 2$, observe that the rate of $\mathcal{C}_1^2$ is $R_2 = \frac{k - v_2}{n_1}$ and the rate of $\mathcal{A}_2^2$ is $R_2 = \frac{v_2}{n_2}$, so we have $(n_1 + n_2) R_2 = k$. Similarly, for $3 \leqslant i \leqslant M$, from the rates of codes $\mathcal{C}_1^i, \mathcal{A}_2^i, \cdots, \mathcal{A}_i^i$, we have $(n_1 + n_2 + \cdots + n_i) R_i = k$. Thus, we prove the rates.

For the decoding error, we prove it for $\mathcal{C}_M$, since the proof also works for any $\mathcal{C}_i$, $1 \leqslant i \leqslant M - 1$. For $\mathcal{C}_M$ over channel $W_M$, we use Algorithm 3 for decoding, where each component decoder is a nearest-codeword decoder as in Algorithm 2. The decoding consists of $M$ steps, so it will succeed if each step is successful. Thus, we can bound the decoding error probability $P_e^{(N_M)}(\mathcal{C}_M)$ by the decoding error probability of each step as

$$P_e^{(N_M)}(\mathcal{C}_M) \leqslant 1 - \left(1 - P_e^{(n_1)}(\mathcal{C}_1^M)\right) \prod_{i=2}^{M} \left(1 - P_e^{(n_i)}(\mathcal{A}_i^M)\right)$$
$$= 1 - \left(1 - P_e^{(\phi_1 N_M)}(\mathcal{C}_1^M)\right) \prod_{i=2}^{M} \left(1 - P_e^{(\phi_i N_M)}(\mathcal{A}_i^M)\right) \quad (9)$$

where constants $\phi_1 = \frac{R_M}{R_1}$ and $\phi_i = \frac{(R_{i-1} - R_i) R_M}{R_i R_{i-1}}$ for $2 \leqslant i \leqslant M$. From the chosen component codes, we already have $P_e^{(\phi_1 N_M)}(\mathcal{C}_1^M) \to 0$ and $P_e^{(\phi_i N_M)}(\mathcal{A}_i^M) \to 0$ as $N_M$ goes to infinity, so in (9), $P_e^{(N_M)}(\mathcal{C}_M) \to 0$ as $N_M$ goes to infinity. Thus, we conclude $\mathcal{C}_M$ can achieve the capacity of $W_M$.

---

**Algorithm 3: Decoding $\mathcal{C}_M$ Over Channel $W_M$**

---

**Input:** received word $\boldsymbol{y} = (\boldsymbol{y}_1, \boldsymbol{y}_2, \cdots, \boldsymbol{y}_M)$.
**Output:** codeword $\boldsymbol{c}_M \in \mathcal{C}_M$.

1: **for** $i = M, M-1, \ldots, 2$ **do**
2:   Let the syndrome $\Lambda_i^i = \boldsymbol{0}$.
3:   $\hat{\boldsymbol{a}}_i^i = \mathscr{D}_{\mathcal{A}_i^M}\left(\boldsymbol{y}_i, (\Lambda_i^i, \Lambda_i^{i+1}, \cdots, \Lambda_i^M)\right)$.
4:   $(\boldsymbol{s}_i, \Lambda_2^i, \cdots, \Lambda_{i-1}^i) = \mathcal{E}_{\mathcal{A}_i^i}^{-1}(\hat{\boldsymbol{a}}_i^i)$.
5: **end for**
6: Let the syndrome $\boldsymbol{s}_1 = \boldsymbol{0}$.
7: $\boldsymbol{c}_1 = \mathscr{D}_{\mathcal{C}_1^M}\left(\boldsymbol{y}_1, (\boldsymbol{s}_1, \boldsymbol{s}_2, \cdots, \boldsymbol{s}_M)\right)$.
8: Let $\boldsymbol{u} = \mathcal{E}_{\mathcal{C}_1}^{-1}(\boldsymbol{c}_1)$, and encode $\boldsymbol{u}$ using Construction 1 to obtain $\boldsymbol{c}_M \in \mathcal{C}_M$.

---

■

## APPENDIX H
## PROOF OF THE EQUIVALENCE BETWEEN THE SYNDROME AND THE FROZEN BITS

*Proof:* We prove the two-level case. Extension to the $M$-level case can be done in a similar way, so is omitted.

We first present the construction in [8] to construct two-level rate-compatible codes $\mathcal{C}_1 \prec \mathcal{C}_2$. Consider two nested polar codes $\mathcal{C}_1^2 = [n_1, n_1 - v_1 - v_2]_2 \subset \mathcal{C}_1^1 = [n_1, k = n_1 - v_1]_2$. The set of frozen bit indices of $\mathcal{C}_1^i$ is denoted by $\mathcal{I}_1^i$ for $i = 1, 2$. It is clear that $|\mathcal{I}_1^1| = v_1$ and $|\mathcal{I}_1^2| = v_1 + v_2$. The nested property of polar codes gives $\mathcal{I}_1^1 \subset \mathcal{I}_1^2$ [9].

For the first step, let a length-$n_1$ vector $\boldsymbol{u}$ have $k$ information bits on the coordinates in $[n_1] \setminus \mathcal{I}_1^1$ and value $0$ on the coordinates in $\mathcal{I}_1^1$. A codeword $\boldsymbol{c}_1 \in \mathcal{C}_1$ is obtained by $\boldsymbol{c}_1 = \boldsymbol{u} G_{n_1}$. Here, the matrix $G_{n_1}$ is $G_{n_1} = B_{n_1} G_2^{\otimes m}$, where $G_2 = \begin{bmatrix} 1 & 0 \\ 1 & 1 \end{bmatrix}$ and $B_{n_1}$ is a bit-reversal permutation matrix defined in [2]. It is known $G_{n_1} = G_{n_1}^{-1}$, i.e., $G_{n_1} G_{n_1} = I$ [5].

For the second step, to get a codeword $\boldsymbol{c}_2 \in \mathcal{C}_2$, they use an auxiliary code $\mathcal{A}_2^2$ to encode the bits on the coordinates in $\mathcal{I}_1^2 \setminus \mathcal{I}_1^1$ of $\boldsymbol{u}$. These bits are denoted by $\pi_{\mathcal{I}_1^2 \setminus \mathcal{I}_1^1}(\boldsymbol{u})$, which will be treated as *frozen bits* during the last step of decoding the code $\mathcal{C}_2$.

Now, for our construction, let us first define the parity-check matrices of $\mathcal{C}_1^1$ and $\mathcal{C}_1^2$ to be $H_{\mathcal{C}_1^1}$ and $H_{\mathcal{C}_1^2} = \begin{bmatrix} H_{\mathcal{C}_1^1} \\ H_{\mathcal{C}_1^2 \setminus \mathcal{C}_1^1} \end{bmatrix}$. Based on Lemma 1 in [5], we have $H_{\mathcal{C}_1^2 \setminus \mathcal{C}_1^1} = P H'_{\mathcal{C}_1^2 \setminus \mathcal{C}_1^1}$, where $H'_{\mathcal{C}_1^2 \setminus \mathcal{C}_1^1}$ is formed by the columns of $G_{n_1}$ with indices in $\mathcal{I}_1^2 \setminus \mathcal{I}_1^1$ and $P$ is a full rank matrix.

The first step of our construction is the same as that in the construction in [8] introduced above. In the second step, we use the same auxiliary code $\mathcal{A}_2^2$ to encode syndrome $\boldsymbol{s}_2$, which is $\boldsymbol{s}_2 = \boldsymbol{c}_1 H_{\mathcal{C}_1^2 \setminus \mathcal{C}_1^1}^T$. In the following, we will prove that $\pi_{\mathcal{I}_1^2 \setminus \mathcal{I}_1^1}(\boldsymbol{u})$ can be one-to-one mapped to the syndrome $\boldsymbol{s}_2$. Specifically, we will show that $\boldsymbol{s}_2 = \pi_{\mathcal{I}_1^2 \setminus \mathcal{I}_1^1}(\boldsymbol{u}) P^T$. To see this, we have the following equations,

$$\boldsymbol{s}_2 = \boldsymbol{c}_1 H_{\mathcal{C}_1^2 \setminus \mathcal{C}_1^1}^T$$
$$= \boldsymbol{u} G_{n_1} H_{\mathcal{C}_1^2 \setminus \mathcal{C}_1^1}^T$$
$$= \pi_{\mathcal{I}_1^2 \setminus \mathcal{I}_1^1}(\boldsymbol{u}) G'_{n_1} H'^T_{\mathcal{C}_1^2 \setminus \mathcal{C}_1^1} P^T$$
$$= \pi_{\mathcal{I}_1^2 \setminus \mathcal{I}_1^1}(\boldsymbol{u}) P^T,$$

where $G'_{n_1}$ is a submatrix of $G_{n_1}$ by taking the rows of $G_{n_1}$ with indices in $\mathcal{I}_1^2 \setminus \mathcal{I}_1^1$. The product $G'_{n_1} H'^T_{\mathcal{C}_1^2 \setminus \mathcal{C}_1^1}$ is an identity matrix $I$, because $H'_{\mathcal{C}_1^2 \setminus \mathcal{C}_1^1}$ is formed by the columns of $G_{n_1}$ with indices in the set $\mathcal{I}_1^2 \setminus \mathcal{I}_1^1$ and also we have $G_{n_1} G_{n_1} = I$. In particular, if we choose $P = I$, then $\boldsymbol{s}_2 = \pi_{\mathcal{I}_1^2 \setminus \mathcal{I}_1^1}(\boldsymbol{u})$. ■